# 40-GHz pulse source based on XPM-induced focusing in normally dispersive optical fibers

J. Nuño, M. Gilles, M. Guasoni, B. Kibler, C. Finot, and J. Fatome*

*Laboratoire Interdisciplinaire Carnot de Bourgogne, UMR 6303 CNRS - Université Bourgogne Franche-Comté
, 9 avenue Alain Savary, 21078 Dijon, France.*

* *jfatome@u-bourgogne.fr*

**We theoretically and experimentally investigate the design of a high-repetition rate source delivering well-separated optical pulses due to the nonlinear compression of a dual-frequency beat signal within a cavity-less normally dispersive fiber-based setup. This system is well described by a set of two coupled nonlinear Schrödinger equations for which the traditional normally dispersive defocusing regime is turned in a focusing temporal lens through a degenerated cross-phase modulation process (XPM). More precisely, the temporal compression of the initial beating is performed by the combined effects of normal dispersion and XPM-induced nonlinear phase shift yield by an intense beat-signal on its weak out-of-phase replica co-propagating with orthogonal polarizations. This adiabatic reshaping process allows us to experimentally demonstrate the generation of a 40-GHz well-separated 3.3-ps pulse train at 1550 nm in a 5-km long normally dispersive fiber.**

The ability to design optical pulse sources at repetition rates of tens of GHz is of a high interest in many fields of photonics including optical communications, sampling, metrology, clocking, sensing, spectral comb or arbitrary waveform generation. In order to develop alternative solutions to traditional mode-locked fiber lasers, numerous cavity-less scenarios have been investigated based on the nonlinear reshaping within optical fibers of an initial beat signal into a train of well-separated pulses [1]. Basically, the nonlinear temporal compression of the initial beating is induced through the focusing regime of the nonlinear Schrödinger equation (NLS), taking advantage of the interplay between the nonlinear Kerr effect and the anomalous dispersive regime [2]. This particular technique has been demonstrated in a wide range of fiber arrangements to produce pedestal-free pulse trains at various repetition rates, ranging from GHz to several THz [1-16]. To this aim, various fiber concatenation systems have been implemented including, dispersion decreasing fibers [3], comb-like or step like fiber profiles [4,17], modulation instability [6], cross-phase modulation (XPM) [7], multiple four-wave mixing [8-13], parametric amplification [15] as well as Raman amplification [16].

Nonetheless, it is noteworthy that even though various techniques of nonlinear compression of a beat signal have been reported in anomalous dispersive optical fibers, only a few exist for normally dispersive fibers. For instance, in two-stage techniques, it is known that an incident optical pulse can be first chirped through self-phase modulation, cross-phase modulation or through an opto-electronic phase modulator and then subsequently compressed by means of a dispersive element inducing an opposite sign of chirp such as a grating or a suitably designed fiber segment [17-19]. However, a less exploited process combining both the nonlinear reshaping and chirp compensation stages can be provided in the defocusing regime of the NLS equation by means of a degenerate cross-phase modulation phenomenon [20-24]. Indeed, it was shown that a XPM-induced focusing of a probe beam can occur in the defocusing regime when it co-propagates orthogonally polarized and bounded between two intense pump pulses travelling at the same group velocity [20-22]. In this configuration, in analogy to the soliton-effect compression technique, the evolution of the dark structure generated in the center of these twin pump pulses induces a XPM-based temporal compression even in normally dispersive fibers, thus creating an adiabatic compressor for the probe beam.

In this letter, we propose to exploit this nonlinear process in order to design a 40-GHz picosecond fiber pulse source centered at 1550 nm. More precisely, we have experimentally demonstrated the nonlinear reshaping of an initial 40-GHz beat signal into a train of well-separated pulses in a normally dispersive fiber thanks to the strong XPM-induced focusing effect generated by its orthogonally polarized high-power out-of-phase replica.

The principle of operation is schematically depicted in Fig. 1. The basic configuration consists in an initial sinusoidal signal co-propagating in a normally dispersive medium with an intense orthogonally polarized out-of-phase pump replica. As shown in Fig. 1a, whilst the defocusing regime induced a nonlinear reshaping of the intense replica into subsequently parabolic then broad and sharp square pulses, it also progressively closes a singularity at its null point characterized by steeper and steeper edges [25-26]. Simultaneously, the XPM induced by this nonlinear dark structure combined with the group velocity dispersion (GVD) of the fiber then turns out the normal dispersive regime into a focusing dynamics for the weak interleaved orthogonal replica. Indeed, as displayed in Fig. 1b by means of a spectro-temporal approach, the pump-induced XPM frequency chirp leads to a blue-shift on the leading-part of the signal whist the trailing edge undergoes a red-shift. Owing to the normal GVD, since blue light travels slower than red, the pulse is then forced to progressively compress as the chirp profile is increasingly steeper. In fact, as a chain of particles trapped in a gradually collapsing periodic potential, here the energy contained in the weak interleaved component is confined through the XPM process. The weak replica is increasingly bounded by this piston effect and is then forced to temporally compress along the fiber length, thus reshaping the initial beating into a train of well-separated pulses.

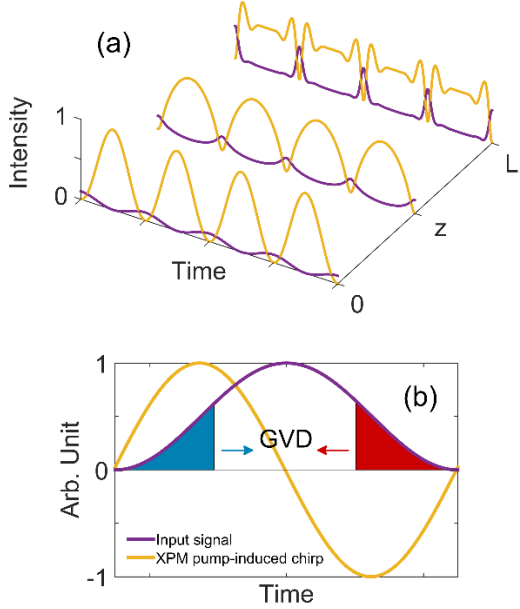

Fig. 1. Principle of operation (a) Evolution of the temporal intensity profiles of orthogonally polarized pump and signal waves as a function of propagation distance (b) Intensity profile of the input signal and XPM-chirp profile induced by the pump on the signal wave.

It is important to notice that, at the opposite of the two-stage techniques for which the signal is nonlinearly chirped in a first segment and then compressed by means of a dispersive element; here the signal simultaneously undergoes both effects in a unique segment, so that the fiber acts as a distributed compressor for the signal wave. This adiabatic dynamics should in principle provide better quality pulses and higher compression factors.

In order to experimentally demonstrate the nonlinear compression of a 40-GHz sinusoidal signal into a train of well-separated pulses in a normally dispersive fiber, we have implemented the experimental setup depicted in Fig. 2. An external cavity laser emitting at 1550 nm is first intensity modulated by means of an intensity modulator (IM). A 20-GHz frequency clock is used to drive the IM at its null-transmission point so as to generate a pure carrier-suppress 40-GHz beating. Subsequently, the resulting beat-signal is then phase modulated thanks to a phase modulator (PM) in order to push back the Brillouin threshold beyond the power levels involved in our experiment. The resulting signal is then injected into a polarization waveshaper from *finisar* which, combined to a polarization-beam splitter (PBS), allows us to successively duplicate, interleave with a fine half-cycle delay of 12.5 ps and polarization multiplex both orthogonal signal replicas. Note that during the polarization multiplexing operation, the power ratio between the pump beam and the orthogonal probe signal is adjusted to 11.5 dB. Both signals are then amplified by means of a high-power Erbium doped fiber amplifier (EDFA) and injected into a 5-km long fiber characterized by a chromatic dispersion $D = -2.5$ ps/nm/km at 1550 nm (second order dispersion $\beta_2 = 3.2$ ps$^2$km$^{-1}$), an attenuation of 0.2 dB/km and a nonlinear Kerr coefficient $\gamma = 1.7$ W$^{-1}$km$^{-1}$. At the output of the system, the resulting 40-GHz pump and signal waves are polarization demultiplexed thanks to a second PBS and characterized in the time domain by means of an optical sampling oscilloscope (OSO, eye-checker from *Alnair Labs*) and in the spectral domain thanks to an optical spectrum analyzer (OSA).

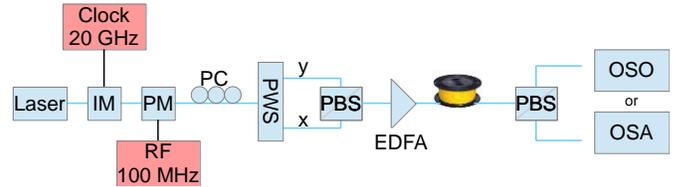

Fig. 2. Experimental setup. IM: intensity modulator, PM: phase modulator, PC: polarization controller, PWS: polarization wave shaper, PBS: polarization beam splitter, EDFA: Erbium doped fiber amplifier, OSO: optical sampling oscilloscope and OSA: optical spectrum analyzer.

The performance of the process is first characterized as a function of the injected power. Fig. 3 summarizes these results and depicts the intensity profile recorded at the output of the fiber for both 40-GHz interleaved signal (Fig. 3a) and pump waves (Fig. 3b) as a function of the input power.

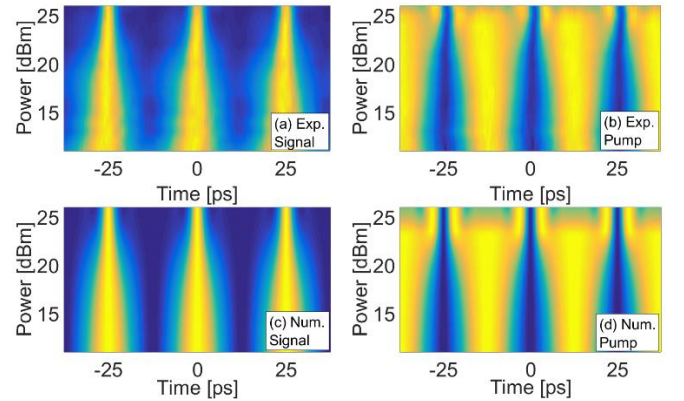

Fig. 3. Output intensity profile as a function of the input power launched into the fiber. Experimental results for the 40-GHz probe (a) and pump (b) signals, respectively (c) & (d) Corresponding numerical simulations.

In the linear regime, when the power launched into the fiber is lower than 12 dBm, the sinusoidal shape of both the signal and pump beams is still preserved. However, as the power is further increased, we can now clearly observe in Fig. 3b the effect of the normally dispersive defocusing regime, which induces a nonlinear reshaping of the intense 40-GHz sinusoidal pump beam into subsequently parabolic and expanding square pulses [25-26]. More importantly, this reshaping is also accompanied by the narrowing of a dark structure around the null points of the pump wave. In parallel, in Fig. 3a, we can clearly observe the effect of XPM on the weak interleaved orthogonal signal which undergoes a focusing dynamics, thus reshaping the initial beat-signal into a train of well-separated short pulses. In analogy to the spatial domain, here the pump beam acts as a positive temporal lens, and hence induces a focusing of the probe signal [22]. The maximum of temporal compression is reached for an injected power of 26 dBm for which the full width at half maximum of the 40-GHz pulses is decreased to 3.3 ps. The last measurements represented in these figures correspond to 26 dBm, for which the Brillouin backscattering begins to be a severe limitation for the process under study.

We compare these experimental results to numerical simulations (Figs. 3c-d) of two orthogonally polarized optical waves propagating in a normally dispersive randomly birefringent standard fiber. The evolution of the complex slowly varying amplitudes of the signal $u$ and pump $v$ waves can be therefore described by the following set of two coupled NLS equations, corresponding to a simplified Manakov model for which nonlinear terms induced by the weak probe have been neglected [27]:

$$i\frac{\partial u}{\partial z} + \frac{\beta_2}{2}\frac{\partial u}{\partial t} + \frac{8}{9}\gamma |v|^2 u = 0$$
$$i\frac{\partial v}{\partial z} + \frac{\beta_2}{2}\frac{\partial v}{\partial t} + \frac{8}{9}\gamma |v|^2 v = 0$$
(1)

where $z$ and $t$ denote the propagation distance and time coordinates. The factor 8/9 takes into account for random variations of the intrinsic birefringence encountered along the fiber length [27]. We can see in Fig. 3c-d a good agreement between the experimental and numerical mapping of the process, thus validating the present technique and its Manakov modeling as well as the assumption that third-order dispersion, losses or Raman effects play a minor role in the nonlinear dynamics and can therefore be neglected.

In order to benefit from design-rules and to avoid the full numerical resolution of Eqs. (1), we have also implemented the following theoretical model. The input pump and signal waves can be written as $v(0,t) = (P/2)^{1/2}\exp(i\omega_1 t) + (P/2)^{1/2}\exp(-i\omega_1 t)$ and $u(0,t) = -i(S/2)^{1/2}\exp(i\omega_1 t) + i(S/2)^{1/2}\exp(-i\omega_1 t)$, where $P$ and $S$ represent the pump and signal power, respectively. The four-wave mixing interaction among the frequency components centered at $\pm\omega_1$ leads to the generation of new spectral components centered at odd-multiples of $\omega_1$ both in the pump and in the signal spectrum. Here we neglect the generation of new components in the pump spectrum, which is an accurate approximation whenever the fiber length is lower than few nonlinear lengths (here less than 5). Indeed, for power levels involved experimentally, first sidebands in the pump spectrum always remain 10-dB bellow the central components. This approximation allows us to write $|v(z,t)|^2$ in a simple form, namely $|v(z,t)|^2 = |v(0,t)|^2 = P + (P/2)\exp(i2\omega_1 t) + (P/2)\exp(-i2\omega_1 t)$. On the other hand, new frequency components cannot be neglected in the case of the weak input signal. Therefore we write $u(z,t)$ as a linear combination of components centered at odd-multiples of $\omega_1$, that is $u(z,t) = \sum s_n(z)\exp(i\omega_n t)$, being $\omega_n = n.\omega_1$ and $n$ an odd-number. If we now insert the aforementioned ansatzes for $|v(z,t)|^2$ and $u(z,t)$ in Eqs.(1), we get the following system of linear differential equations (LDE) describing the spatial dynamics of $s_n(z)$: $\partial s_n/\partial z = ia_n s_n + ib(s_{n+2} + s_{n-2})$, where the constant coefficients $a_n$ and $b$ read as $a_n = (n^2\omega_1^2\beta_2/2 + 8/9\gamma P)$ and $b = 4/9\gamma P$. This system of LDE can then be easily solved when it is truncated at a finite number $n=n_c$ and by considering the appropriate initial conditions $s_1(0) = -i(S/2)^{1/2}$, $s_{-1}(0) = i(S/2)^{1/2}$ and $s_n(0)=0$ for $n \neq \pm 1$, which permits us to estimate the signal at the fiber output as well as its compression ratio. Typically, the larger $n_c$ the more accurate is the estimation of $v(z,t)$, but in practice $n_c = 7$ (8 components) is sufficient to get a good estimation.

In order to better evaluate the quality of the generated pulses, we have displayed in Fig. 4a the temporal intensity profile of the compressed signal for a total injected power of 26 dBm. Here, the experimental measurement (solid line) is compared with the results obtained from the numerical solving of Eqs. (1) (diamonds) as well as the analytical solution obtained from the procedure described above (circles).

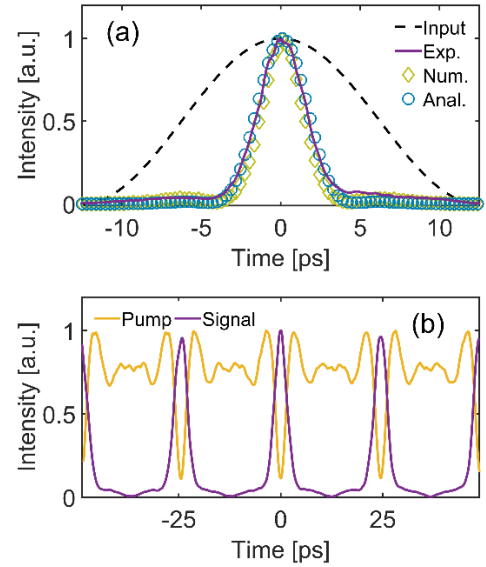

**Fig. 4.** (a) Output intensity profile of the 40-GHz compressed signal for an input power of 26 dBm. Experimental results (purple solid-line) are compared with numerical simulations (green diamonds) and analytic model (blue circles) (b) Output intensity profiles of both signal and pump waves recorded at the output of the fiber for an injected power of 26 dBm.

The resulting intensity profile shows 3.3-ps well separated pulses, with small residual pedestal. We can observed a good agreement between all the curves, thus allowing an efficient prediction and design of this optical pulse source. In Fig. 4b, the temporal output experimental profiles of both the pump and signal waves are simultaneously displayed and confirm the complementary evolution undergone by the two orthogonal components. Note the perfect temporal interleaving between the pump and signal waves with a fix delay of 12.5 ps. Indeed, if the initial signals are not perfectly out-of-phase, the nonlinear focusing will be less effective with the appearance of an asymmetry in the resulting temporal profile.

Fig. 5a shows the evolution of the compression ratio as a function of injected power. Experimental measurements (stars) are compared to theoretical predictions (circles) and numerical simulations (diamonds). The compression factor is exponentially growing and the pulse are compressed approximately 4 times fold when the input power reaches 26 dBm. Note that beyond 26 dBm, the experimental compression ratio begins to saturate due to the impairments induced by the Brillouin backscattering.

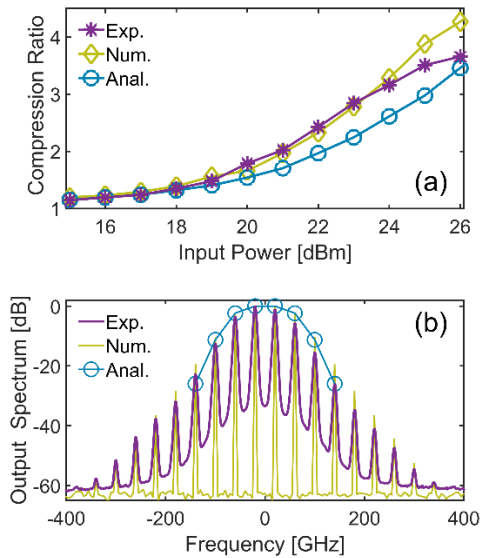

**Fig. 5.** (a) Evolution of the compression ratio as a function of the input power: Experimental results (purple stars), numerical simulations (green diamonds) and theoretical predictions (blue circles) (b) Output spectrum of the signal wave for an injected power of 26 dBm. Measurements (purple solid-line) are compared with numerical simulations (green) and analytic model (blue circles).

Finally, moving to the spectral domain, the phenomenon can be characterized as the generation of new frequency components related to the time-dependence of the nonlinear phase shift induced by XPM, as illustrated in Fig. 5b for a 26-dBm input power. Although numerous components have been generated experimentally, only eight have to be considered for the analytical description (circles) as the power difference between the inner and out-of-band components is larger than 20 dB. These results also confirm once again the quantitative agreement between theory (circles), simulations (green) and experiments (purple).

In summary, we have proposed and experimentally demonstrated a novel technique for the generation of high repetition rate picosecond pulses. The present system is based on the nonlinear reshaping of an initial beat signal within a normally dispersive optical fiber for which the traditional defocusing regime is turned in a focusing temporal lens through a degenerated cross-phase modulation process. More precisely, the focusing of the beat-signal is here induced by the combined effects of normal dispersion and nonlinear phase shift generated by means of a co-propagating interleaved and orthogonally polarized intense replica. We have successfully implemented the compression of a 40-GHz sinusoidal beating centered at 1550 nm in a well-separated 3.3-ps pulse train within a 5-km long normally dispersive fiber. Numerical and theoretical results are in good agreement with our experimental recordings. Finally, this XPM-trapping effect could find more applications in photonics, such as the resynchronization of signals as well as clocking or sampling operations.

This research is funded by the European Research Council under Grant Agreement 306633, ERC PETAL (www.facebook.com/petal.inside). We also acknowledge the financial support from the Conseil Régional de Bourgogne, the FEDER and Labex ACTION (ANR-11-LABX-0001-01).